\documentclass[longauth,letter,a4paper,traditabstract]{aa}  
\usepackage{graphicx,amsmath,txfonts,natbib,astro}
\usepackage[version=3]{mhchem}
\usepackage[normalem]{ulem}
\begin{document}
\def\iras{IRAS16293-2422}
\def\ncrit{\ensuremath{n_{\rm crit}}}
\def\tc {\ensuremath{T_C}}
\def\ntot{\ensuremath{N_{\rm tot}}}
\def\aul{\ensuremath{A_{ul}}}
\def\gu{\ensuremath{g_{u}}}
\def\tauk{\ensuremath{\tau_{ul}}}
\def\tant{\ensuremath{{T_{\rm ant}}}}
\def\jnu{\ensuremath{J_\nu}}
\def\tcmb{\ensuremath{T_{\rm CMB}}}
\def\jnuex{\ensuremath{{\tilde{T}_{\rm ex}}}}
\def\jnucmb{\ensuremath{{\tilde{T}_{\rm cmb}}}}
\def\osu{\texttt{osu.09.2008}}
\def\tfm#1{\tablefootmark{#1}}
\def\tft#1#2{{\tablefoottext{#1}{#2}}}
\def\tfoot#1{{\tablefoot{\scriptsize #1}}}
\def\ratio{\ce{NH}:\ce{NH2}:\ce{NH3}}
\def\corr#1{\textbf{#1}}
\title{Nitrogen hydrides in the cold envelope of
  IRAS16293-2422\thanks{Herschel is an ESA space observatory with
    science instruments provided by European-led principal
    Investigator consortia and with important participation from
    NASA.}}

   \titlerunning{Nitrogen hydrides in the envelope of IRAS16293-2422}

   \author{%
P.~Hily-Blant	\inst{1}	\and
S.~Maret	\inst{1}	\and
A.~Bacmann	\inst{1,2}	\and
S.~Bottinelli	\inst{6}        \and
B.~Parise	\inst{10}	\and
E.~Caux	        \inst{6}	\and
A.~Faure        \inst{1}	\and\\
E.A.~Bergin	\inst{25}	\and
G.A.~Blake	\inst{3}	\and
A.~Castets	\inst{1}	\and
C.~Ceccarelli	\inst{1}	\and
J.~Cernicharo	\inst{9}	\and
A.~Coutens	\inst{6}	\and
N.~Crimier	\inst{1,9}	\and
K.~Demyk	\inst{6}	\and
C.~Dominik	\inst{12,13}	\and
M.~Gerin	\inst{28}	\and
P.~Hennebelle	\inst{28}	\and
T.~Henning	\inst{26}	\and
C.~Kahane	\inst{1}	\and
A.~Klotz	\inst{6}	\and
G.~Melnick	\inst{18}	\and
L.~Pagani	\inst{8}	\and
P.~Schilke	\inst{10,20}	\and
C.~Vastel	\inst{6}	\and
V.~Wakelam	\inst{2}	\and
A.~Walters	\inst{6}	\and
A.~Baudry	\inst{2}	\and
T.~Bell	        \inst{3}	\and
M.~Benedettini	\inst{4}	\and
A.~Boogert	\inst{5}	\and
S.~Cabrit	\inst{8}	\and
P.~Caselli	\inst{7}	\and
C.~Codella	\inst{11}	\and
C.~Comito	\inst{10}	\and
P.~Encrenaz	\inst{8}	\and
E.~Falgarone	\inst{28}	\and
A.~Fuente	\inst{14}	\and
P.F.~Goldsmith	\inst{15}	\and
F.~Helmich	\inst{16}	\and
E.~Herbst	\inst{17}	\and
T.~Jacq	        \inst{2}	\and
M.~Kama	        \inst{12}	\and
W.~Langer	\inst{15}	\and
B.~Lefloch	\inst{1}	\and
D.~Lis	        \inst{3}	\and
S.~Lord	        \inst{5}	\and
A.~Lorenzani	\inst{11}	\and
D.~Neufeld	\inst{19}	\and
B.~Nisini	\inst{24}	\and
S.~Pacheco	\inst{1}	\and
T.~Phillips	\inst{3}	\and
M.~Salez	\inst{8}	\and
P.~Saraceno	\inst{4}	\and
K.~Schuster	\inst{21}	\and
X.~Tielens	\inst{22}	\and
F.~van~der~Tak  \inst{16,27}    \and
M.H.D.~van~der~Wiel \inst{16,27} \and	
S.~Viti	        \inst{23}	\and
F.~Wyrowski	\inst{10}	\and
H.~Yorke	\inst{15}
}

\institute{		
$^1$
Laboratoire d'Astrophysique de Grenoble, UMR 5571-CNRS, Universit\'e Joseph Fourier, Grenoble, France\\
$^2$
Universit\'{e} de Bordeaux, Laboratoire d'Astrophysique de Bordeaux, CNRS/INSU, UMR 5804, Floirac, France\\
$^3$
California Institute of Technology, Pasadena, USA\\
$^4$
INAF - Istituto di Fisica dello Spazio Interplanetario, Roma, Italy\\
$^5$
Infared Processing and Analysis Center,  Caltech, Pasadena, USA\\
$^6$
Centre d'Etude Spatiale des Rayonnements, Universit\'e Paul Sabatier, Toulouse 3, CNRS UMR 5187, Toulouse, France\\
$^7$
School of Physics and Astronomy, University of Leeds, Leeds UK\\
$^8$
LERMA and UMR 8112 du CNRS, Observatoire de Paris, 61 Av. de l'Observatoire, 75014 Paris, France\\
$^9$
Centro de Astrobiolog\'{\i}a, CSIC-INTA, Madrid, Spain\\
$^{10}$
Max-Planck-Institut f\"{u}r Radioastronomie, Bonn, Germany\\
$^{11}$
INAF Osservatorio Astrofisico di Arcetri, Florence Italy\\
$^{12}$
Astronomical Institute 'Anton Pannekoek', University of Amsterdam, Amsterdam, The Netherlands\\
$^{13}$
Department of Astrophysics/IMAPP, Radboud University Nijmegen,  Nijmegen, The Netherlands\\
$^{14}$
IGN Observatorio Astron\'{o}mico Nacional, Alcal\'{a} de Henares, Spain\\
$^{15}$
Jet Propulsion Laboratory,  Caltech, Pasadena, CA 91109, USA\\
$^{16}$
SRON Netherlands Institute for Space Research, Groningen, The Netherlands\\
$^{17}$
Ohio State University, Columbus, OH, USA\\
$^{18}$
Harvard-Smithsonian Center for Astrophysics, Cambridge MA, USA\\
$^{19}$
Johns Hopkins University, Baltimore MD,  USA\\
$^{20}$
Physikalisches Institut, Universit\"{a}t zu K\"{o}ln, K\"{o}ln, Germany\\
$^{21}$
Institut de RadioAstronomie Millim\'etrique, Grenoble - France\\
$^{22}$
Leiden Observatory, Leiden University, Leiden, The Netherlands\\
$^{23}$
Department of Physics and Astronomy, University College London, London, UK\\
$^{24}$
INAF - Osservatorio Astronomico di Roma, Monte Porzio Catone, Italy\\
$^{25}$
Department of Astronomy, University of Michigan, Ann Arbor, USA\\
$^{26}$
Max-Planck-Institut f\"ur Astronomie, Heidelberg, Germany\\
$^{27}$
Kapteyn Astronomical Institute, University of Groningen, The Netherlands\\
$^{28}$
LERMA, UMR 8112-CNRS, Ecole Normale Sup\'erieure et Observatoire de
Paris, France\\
          }

\abstract{Nitrogen is the fifth most abundant element in the Universe,
  yet the gas-phase chemistry of N-bearing species remains poorly
  understood. Nitrogen hydrides are key molecules of nitrogen
  chemistry. Their abundance ratios place strong constraints on the
  production pathways and reaction rates of nitrogen-bearing
  molecules. We observed the class 0 protostar IRAS16293-2422 with the
  heterodyne instrument HIFI, covering most of the frequency range
  from 0.48 to 1.78~THz at high spectral resolution. The hyperfine
  structure of the amidogen radical o-NH$_2$ is resolved and seen in
  absorption against the continuum of the protostar. Several
  transitions of ammonia from 1.2 to 1.8~THz are also seen in
  absorption. These lines trace the low-density envelope of the
  protostar. Column densities and abundances are estimated for each
  hydride. We find that \ratio$\approx$5:1:300. {Dark clouds chemical
    models predict steady-state abundances of \ce{NH2} and \ce{NH3} in
    reasonable agreement with the present observations, whilst that of
    NH is underpredicted by more than one order of magnitude, even
    using updated kinetic rates. Additional modelling of the nitrogen
    gas-phase chemistry in dark-cloud conditions is necessary before
    having recourse to heterogen processes.}}

\keywords{ISM: abundances, Astrochemistry, ISM individual objects:
  IRAS 16293-2422}

\maketitle


\section{Introduction}

Nitrogen is the fifth most abundant element in the Universe and is a
fundamental component of molecules associated with
life. Nitrogen-bearing molecules are routinely observed towards a wide
variety of environments, from the diffuse interstellar medium
\citep{liszt2001} to pre-stellar cores \citep{bergin2007} and
protoplanetary disks \citep{dutrey1997}. Complex N-bearing molecules
are also observed towards star-forming regions \citep{herbst2009}. The
chemical network of nitrogen is apparently simple in that a small set
of reactions is involved \citep[][hereafter PdF90]{pineau1990}. The
chemistry of nitrogen has been modelled in various environments with
moderate success, including typical dark cloud conditions
\citep{millar1991,lebourlot1991}, shocks (PdF90), pre-stellar cores
\citep{flower2006,maret2006,hilyblant2010n}, and photo-dissociation
regions (PDR) \citep{sternberg1995}. {One major unknown is the total
  abundance, in dense and shielded environments, of gas-phase
  nitrogen, the reservoir of which consist of N and/or
  \ce{N2}. Because they are not directly observable, estimates of
  their abundances rely on observations of other N-bearing compounds
  and chemical modelling. Observational constraints of the dominant
  chemical pathways of the nitrogen chemistry and their kinetic rates
  are thus crucial.}


In this respect, nitrogen hydrides are of utmost importance since they
are {among the first neutral N-bearing molecules formed in an
  initially atomic gas dominated by hydrogen and helium}. Ammonia was
among the first interstellar molecules detected in emission towards
the Galactic centre \citep{cheung1968}. The lightest radical, imidogen
\ce{NH}, was observed in absorption by \cite{meyer1991} along the
diffuse line of sight towards $\zeta$Per. Gas-phase models were found
to underestimate the abundance of NH, and dust grains were then
proposed to solve part of the discrepancy
\citep{meyer1991,wagenblast1993}. Amidogen (\ce{NH2}) was
observed in absorption by \cite{vandishoeck1993} from dense gas in
Sgr~B2.  This source was also targeted by \cite{goicoechea2004} with
ISO. The ratios of the three hydrides {were} found to be
NH:\ce{NH2}:\ce{NH3}$\approx$ 1:10:100, incompatible with the dark
cloud value \ce{NH3}/\ce{NH2}$<3$ predicted by
\cite{millar1991}. Unfortunately, the modelling of the chemistry in
Sgr~B2 is difficult due to the complexity of the source, which probably incorporates shock dynamics. The \ratio\ ratios measured in Sgr~B2 may
thus not be representative of cold dark clouds, and the chemistry of
nitrogen hydrides in these environments remains largely unexplored.


In this paper, we present HIFI observations of the submillimetre lines
of \ce{NH2} and \ce{NH3} in absorption against the continuum of the
class~0 protostar \iras. Section 2 summarizes the observation strategy
and data reduction. In Sect. 3 we derive the column densities of
\ce{NH2} and \ce{NH3}. Abundances of the three lightest nitrogen
hydrides are estimated in Sect. 4, and compared to steady-state
models using updated reaction rates.

\section{Observations and data reduction}

The solar-mass protostar \iras\ was observed with the HIFI instrument
onboard the Herschel Space Observatory, as part of the HIFI guaranteed
time key program CHESS \citep{ceccarelli2010}. Full spectral coverage
of bands 1a ($480-560$\,GHz), 3b ($858-961$\,GHz), 4a
($949-1061$\,GHz), 5a ($1.12-1.24$~THz), and 7a ($1.70-1.79$~THz) were
performed on 2010 Mar 1, 3, and 19, using the Spectral Scan DSB mode
with optimization of the continuum. The Wide Band Spectrometre (WBS)
was used as a backend, providing us with a spectral resolution of
1.1~MHz over an instantaneous bandwidth of $4\times 1$\,GHz. The
targeted coordinates were $\alpha_{2000}$ = 16$^h$ 32$^m$ 22$^s$75,
$\delta_{2000}$ = $-$ 24$\degr$ 28$\arcmin$ 34.2$\arcsec$. The {two}
reference positions were situated approximately 3\arcmin\ east and
west of the source. The beam size is well approximated by
$HPBW=21.5\arcsec/\nu_{\rm THz}$. For the analysis, intensities were
then brought to a main-beam temperature scale using $\feff=0.96$ and
$\beff=0.70$.

The data were processed using the standard HIFI pipeline HIPE 2.8
\citep{ott2010} up to frequency and amplitude calibrations (level 2).
For the SIS bands 1 to 5, a single local-oscillator--tuning spectrum
consists of 4 sub-bands of 1~GHz for each polarization. The 1~GHz
chunks for bands 1 to 5 are then exported as FITS files in the
CLASS90/GILDAS format\footnote{http://www.iram.fr/IRAMFR/GILDAS}
\citep[][]{IRAM_report_2005-1} for subsequent data reduction and
analysis. Despuring and residual bandpass effect subtraction were performed
in CLASS90 using generic spectral-survey tools developed in our
group. Sideband deconvolution is computed with the minimization
algorithm of \cite{comito2002} implemented into CLASS90. Line
identification used the Weeds CLASS90--add-on developed by
\cite{weeds2010}, which provides an efficient interface to the public
CDMS and JPL spectroscopic databases \citep{cdms,jpl}.

\section{Results}

\begin{figure}
  \centering
  \includegraphics[height=1\hsize,angle=-90]{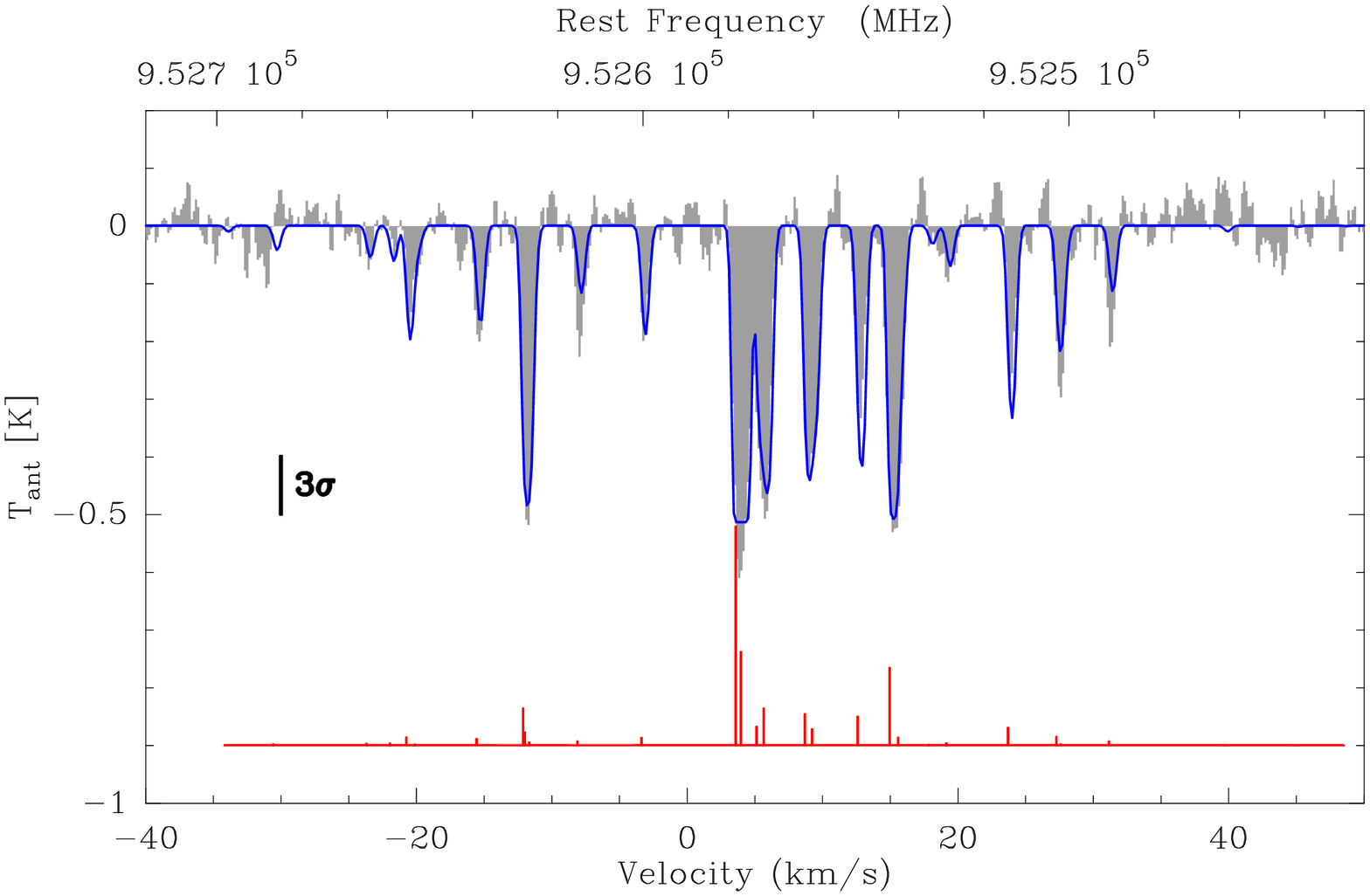}\bigskip\\
  \includegraphics[height=1\hsize,angle=-90]{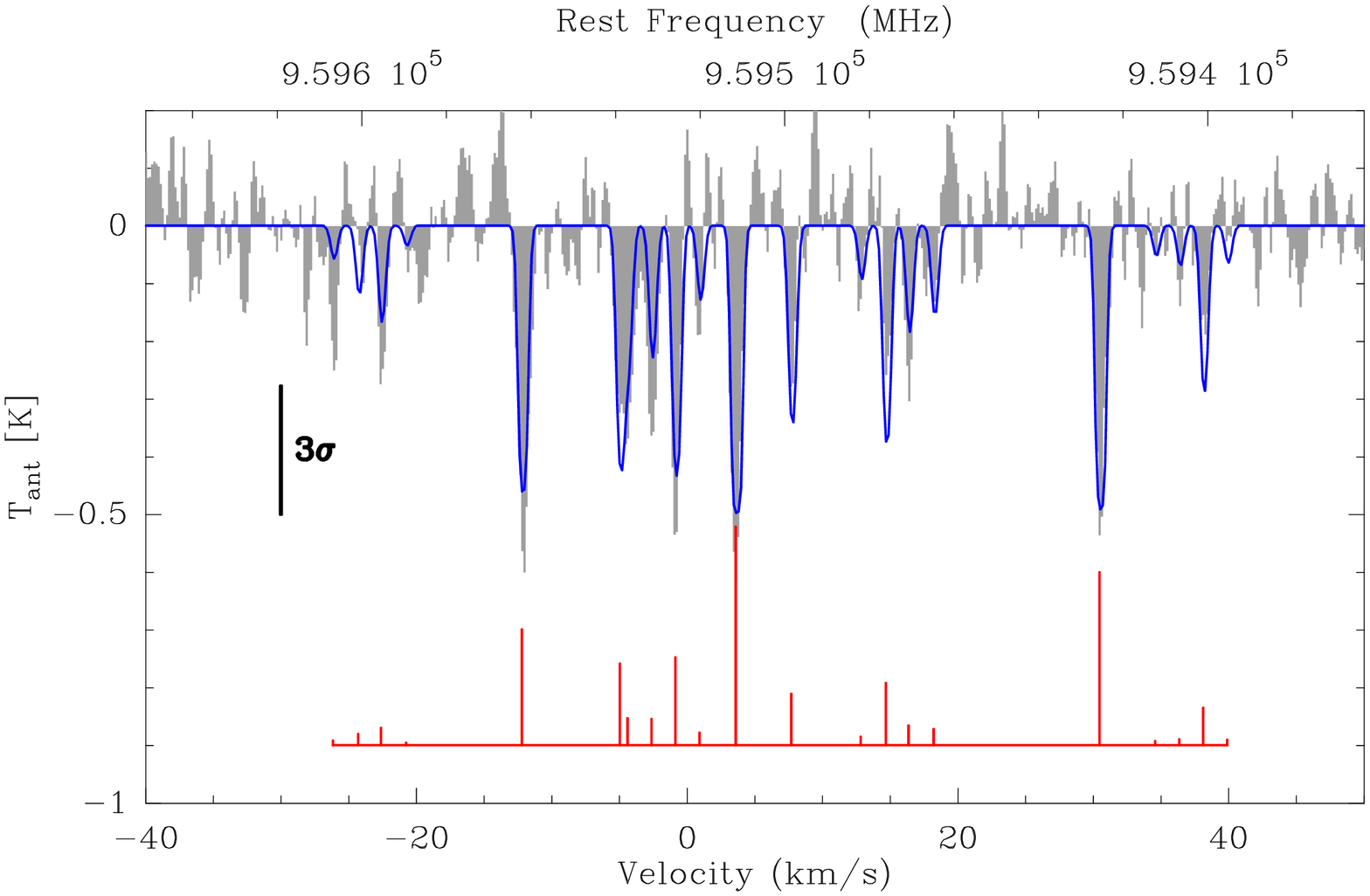}
  \caption{Absorption spectra of o-\ce{NH2} at 952~GHz and 959~GHz,
    with the HFS fit overlaid (line). The relative intensities in the
    optically thin LTE limit are indicated at the bottom.}
  \label{fig:nh2}
\end{figure}

\begin{figure}
  \centering
  \includegraphics[width=0.75\hsize,angle=-90]{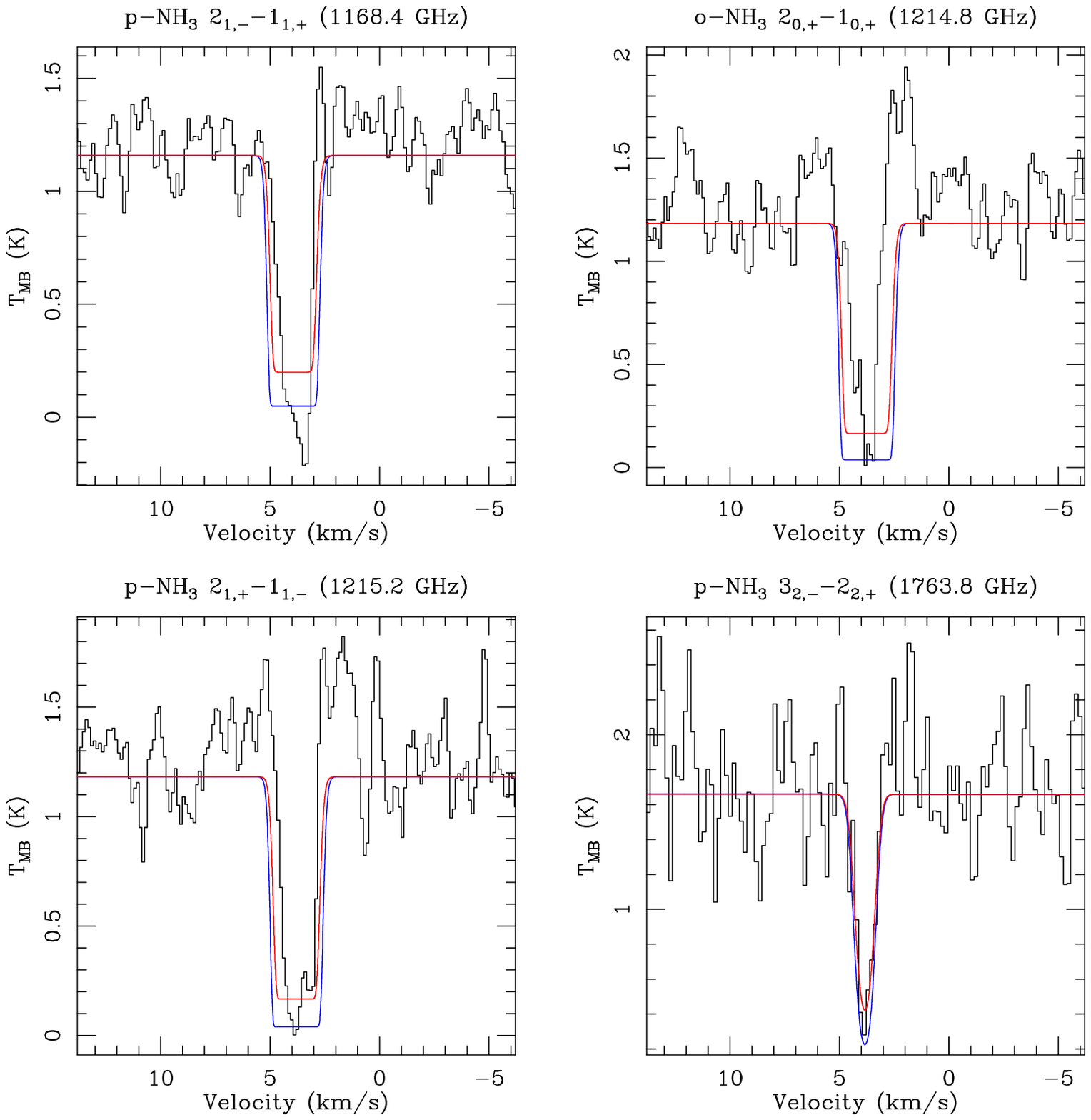}
  \caption{\nhhh\ absorption lines from 1.168 to 1.764~THz. LTE
    predictions are shown in red (\texc=10~K,
    $N(\nhhh)=3.5\tdix{15}$~\cc) and blue ($\texc=8$~K,
    $N(\nhhh)=2\tdix{16}$~\cc).}
  \label{fig:nh3}
\end{figure}

Figure~\ref{fig:nh2} shows the detection of the hyperfine structure
(HFS) of the $N=0-1,\,J=\frac{1}{2}-\frac{3}{2}$, and
$J=\frac{1}{2}-\frac{1}{2}$ transitions of amidogen in its ortho form,
with their strongest components at the rest frequencies 952.578354~GHz
and 959.511716~GHz, respectively \citep{muller1999}. The HFS is almost
entirely resolved with an intensity ratio that clearly deviates from
optically thin LTE excitation. The $3\sigma$ noise levels are
indicated in each panel. The para-\ce{NH2} line was not detected
{and a 5$\sigma$ upper limit to the main HFS component at
  947.725~GHz is $\int \tau \ud v \le
  0.2$~\kms}. Figure~\ref{fig:nh3} shows several transitions of
ammonia seen in absorption, from 1168.4 to 1763.8~GHz. The fundamental
rotational transition at 572.6~GHz was also detected in emission (see
Fig.~\ref{fig:nh3emiss}) but because it contains both emission and absorption,
it is not discussed in this paper.


\begin{table*}
  \centering
  \caption{Column densities of nitrogen hydrides towards \iras.}
  \begin{tabular}{l c c c c c c c c c c c}\hline
    Species & Transition & Component\tfm{a}  & 
    Frequency& HPBW & $T_{C,mb}$\tfm{b} 
    & $T_l$\tfm{c}
    & \tauk & \texc & $FWHM$\tfm{d} & $N_{\rm tot}$ \\
    &            & & GHz  & arcsec & K &K && K & \kms & \dix{14}\cc \\
    \hline
    NH\tfm{e}& $0_1\ra1_0$ & $\frac{3}{2},\frac{5}{2}\ra\frac{1}{2},\frac{3}{2}$
    &  946.476 & 23 & 0.8  &  $-3.8\pm0.6$  & $2.8\pm0.7$ & 9.5 & $0.60$ & $2.20\pm0.80$   \\
    o-\ce{NH2}  & $0_{00\frac{1}{2}}\ra1_{11\frac{3}{2}}$ & $\frac{3}{2},\frac{5}{2}\ra\frac{5}{2},\frac{7}{2}$
    &  952.578 & 23 & 0.9  &  $-9.0\pm0.5$ & $12.8\pm0.7$ & 8.5 & $0.60$ & $0.40\pm0.06$\\ 
    & $0_{00\frac{1}{2}}\ra1_{11\frac{1}{2}}$ & $\frac{3}{2},\frac{5}{2}\ra\frac{3}{2},\frac{5}{2}$
    & 959.512 & 22 & 0.9  &  $-2.5\pm0.2$ & $4.9\pm0.7$ & 9.5 & $0.60$ & $0.59\pm0.12$\\
    p-\nhhh & $1 - 2$ & $(1,-) - (1,-)$  & 1168.453 & 18 & 1.2  & &$300-70$  &  $8-10$ & 0.50 & $200-35$\\
    o-\nhhh & $1 - 2$ & $(0,+) - (0,+)$  & 1214.853 & 18 & 1.3  & &$470-130$ &  $8-10$ & 0.50 & $200-35$\\
    p-\nhhh & $1 - 2$ & $(1,+) - (1,+)$  & 1215.246 & 18 & 1.3  & &$330-80$  &  $8-10$ & 0.50 & $200-35$\\
    p-\nhhh & $2 - 3$ & $(2,-) - (2,-)$  & 1763.823 & 12 & 2.2  & &$2.0-1.4$ &  $8-10$ & 0.50 & $200-35$\\
    \hline
  \end{tabular}
  \label{tab:res}
  \tfoot{ \tft{a}{For NH, the quantum numbers for the rotational
      transition $N_J$ are $\mathbf{F_1=I_H+J}$ and
      $\mathbf{F=I_N+F_1}$ \citep{klaus1997}. For the $N_{K_aK_cJ}$
      rotational transition of \ce{NH2}, the quantum numbers are
      $\mathbf{F_1=I_N+J}$ and $\mathbf{F=I_H+F_1}$
      \citep{muller1999}. In the case of ammonia, quantum numbers are
      given separately for $\mathbf{J=N+S}$ and $(K,\epsilon)$, where
      $\epsilon$ is the symmetry index \citep[see][]{maret2009}. {For
        these lines, the frequency given is that of the brightest HF
        component.}}  \tft{b}{Single-sideband continuum in a
      \tmb\ scale.} \tft{c}{{$T_l =
        \tauk[J_\nu(\texc)-J_\nu(\tcmb)-T_{C,mb}]$ from the HFS
        fit. In the case of ammonia, see Sect.~3.}}  \tft{d}{A
      conservative uncertainty of 0.25~MHz (0.08~\kms) imposed by the
      HIPE 2.8 pipeline was retained.}  \tft{e}{The integrated line
      opacity is calculated as $\int \tau \ud v = 1.06\, FWHM \times
      \tauk$.}  \tft{d}{\cite{bacmann2010}}.}
\end{table*}
%
%
%

All column densities are derived assuming a single excitation
temperature $\texc$ for each molecule.  {The opacity of each HFS
  component and the excitation temperature are determined by
  simultaneously fitting all HFS components, constraining the
  opacities to scale with $\aul\gu$. The fit is performed in CLASS90
  by applying the HFS method to the continuum-subtracted spectra $\tmb
  = [\jnu(\texc)-\jnu(\tcmb)-\tc]\,(1-e^{-\tauk})$. The total column
  density reads $\ntot = ({8\pi\nu^3}/{c^3})\,{Q}/({\aul\gu})\times
  \int \tauk \ud v \times e^{E_l/k\texc}/[1-e^{-h\nu/k\texc}]$, where
  $Q$ is the partition function at \texc. In deriving the excitation
  temperature, we assumed equal filling factors for the absorbing gas
  and the background continuum radiation. The SSB continuum intensity
  $\tc$, needed to derive \texc, is estimated as half the median of
  each 1~GHz chunk computed in line-free spectral windows prior to
  deconvolution, assuming equal gains in the two sidebands (see
  details in Appendix~\ref{sec:cdens}). It is found to increase with
  frequency and, for rest frequencies $\nu_{\rm THz}=0.492$ to
  1.242~THz, \tc\ is well approximated by $\tc (\tant,{\rm [K]}) =
  1.10\,\nu_{\rm THz}-0.42$ (see Fig.~\ref{fig:contlevel}). At higher
  frequencies, $\tc$ is estimated from the deconvolution tool in the
  HIPE software, assuming equal sideband gains.}


For ammonia, the hyperfine components in each lines are not
resolved out by our observations, so the column density and excitation
temperature were determined by simultaneously fitting the 4 lines
shown in Fig.~\ref{fig:nh3}. The 1763.8~GHz line has a well
constrained centre-line opacity, yet the excitation temperature and
column density remain degenerate. We therefore varied the
excitation temperature between 5 to 15~K, and adjusted accordingly the
column density until the observed line profiles were reproduced. The
line width was fixed to 0.5~\kms. For $\texc>10$~K, the absorption
lines at 1.11684, 1.2148, and 1.2152~THz become significantly weaker
than observed. In contrast, for $\texc < 8$~K, the column density
needed to reproduce the 1.7~THz line becomes so large that the 1.2~THz
lines broaden significantly. The full set of constraints points
towards $\texc\approx 8-10$~K and $N_{\rm tot}\approx
2.0\tdix{16}-3.5\tdix{15}$~\cc. Figure~\ref{fig:nh3} shows the results
of the modelled lines corresponding to $\texc=8$ and 10~K. The
non-detection of the 1.7635 and 1.7636 THz absorption lines (not shown
in Fig.~\ref{fig:nh3}) is consistent with these models.

The results are summarized in Table~\ref{tab:res}. The excitation
temperature of the 952 and 959~GHz transitions of o-\ce{NH2} are
significantly higher than 2.73~K. The different values for each HFS
most likely results from the LTE assumption not being entirely
valid. However, $\texc\ll h\nu/k$, so that the corresponding
uncertainties have negligible consequences on the column density.
{The thermalized ortho:para} ratio is expected to be large at low
temperatures ($\approx e^{30.39/T}$, see Fig.~\ref{fig:opnh2}). Our
two determinations of the o-\ce{NH2} column density thus give an
average for the total \ce{NH2} column density. The resulting \ce{NH2}
and \ce{NH3} column densities are $0.44\pm0.07\tdix{14}$~\cc\ and
$\approx35-200\tdix{14}$~\cc, respectively. The column density ratios
are thus $\ce{NH}/\ce{NH2} = 5.0\pm1.2$ and
$\ce{NH3}/\ce{NH2}=90-500$, or NH:\ce{NH2}:\ce{NH3}$\approx$5:1:300.

%

\section{Discussion}

From Table~\ref{tab:res}, the low excitation temperatures indicate
that all lines are sub-thermally excited and therefore most likely trace
regions with densities much lower than their critical densities,
which are of the order of \dix{7}~\ccc. The physical source model of
\cite{crimier2010} predicts densities lower than $\approx
\dix{6}$~\ccc\ for radii larger than 2400~AU or 17\arcsec\ at a
distance of 120~pc, comparable to the $HPBW$ of the present
observations, thus supporting the assumption of equal filling-factors
for the absorbing gas and the continuum emission. At these radii, the
modelled gas temperature is lower than 20~K. {The NH and
  \ce{NH2}} linewidths are thus dominated by non-thermal broadening
suggesting that turbulence has not been dissipated in the absorbing
gas.

To place constraints on the average abundances in the foreground
absorbing material, we need to measure the total H column density, which is not
directly observable. The one-dimensional density profile of
\cite{crimier2010} is extended to lower densities as
$\nhh(r)=3\tdix{8}\, (r/85\au)^{-1.8}+\dix{3}$~\ccc, to allow
for a low-density envelope. The column density {profile},
convolved by the HIFI beam, is dominated by the lines-of-sight close
to the centre. Considering only the gas in the regions with
$\nhh<\dix{6}$~\ccc, the column density is $N(\hh)\approx
8\tdix{22}$~\cc\ or 80 magnitudes of visual extinction (assuming
standard dust properties and that all H is molecular). Were the
hydrides absorption to occur at densities lower than \dix{4} or
\dix{5}~\ccc, the total column density would decrease by a factor
3. Therefore, in the following we estimate the abundances assuming
$N(\hh)=5.5\pm2.5\tdix{22}$~\cc. Results are summarized in
Table~\ref{tab:model}.

The column density ratios observed in the cold envelope of
\iras\ differ from those derived by \cite{goicoechea2004}, who found
NH:\ce{NH2}:\ce{NH3}=1:10:100 towards Sgr~B2. These authors noted that
these ratios are not consistent with typical dark cloud conditions but
can be explained by shock chemistry. Though the \ce{NH3}/\ce{NH2}
ratios are similar in both the cold envelope and the Sgr~B2 region, the
NH/\ce{NH2} ratios are drastically different, suggesting that different
chemistry is at work in the two sources. To study this, we have employed the
\texttt{Astrochem} gas-phase chemical code of \cite{maret2010}
combined with modified versions of the \osu\ chemical
network\footnote{\texttt{http://www.physics.ohio-state.edu/\~{}eric/research.html}}. Typical
physical conditions for a cold molecular cloud (gas temperature
$T=10$~K, $\nh = 2\times 10^4$~\ccc, $A_v=10$~mag) and a cosmic-ray
ionization rate $\zeta=1.3\times 10^{-17}$~\s\ were used. Higher
extinctions would not modify the predicted steady-state abundances.
In a similar way, higher densities would shorten the time to reach
a steady-state but would not alter the corresponding abundances. The
initial abundances are taken from \cite{wakelam2008} {for which
  the C/O gas-phase abundance ratio is 0.41}. The resulting
time-dependent fractional abundances of the nitrogen hydrides are
shown on Fig.~\ref{fig:model}. The steady-state is reached after {a
  few \dix{6}~yrs}, for which the \ratio\ abundance ratios are
0.2:1:190 (see Table~\ref{tab:model}). {We note that the
  \ratio\ ratios are roughly constant for times $> 10^5$ yrs.} If
\ce{NH3}/\ce{NH2} is consistent with the observations to within a
factor of 2, the steady-state NH/\ce{NH2} ratio is too small by more
than one order of magnitude because the abundance of NH is
underpredicted by our model.

{\ce{NH2} and \ce{NH3} are principally formed by the
  dissociative recombination (DR) of \ce{NH4+}
  \citep{lebourlot1991}. The formation of NH is dominated by
  \ce{NH2}(O,OH)NH, unless the NH channel of the DR of \ce{N2H+} has a
  non-zero branching ratio (BR). There are still disparate results for
  the BR of these DR reactions, and the most recent literature
  \citep[see Table~\ref{tab:model}\,and][]{florescu2006} suggest
  uncertainties of at least 10\% in the BR.  We conducted three
  model calculations, where we varied the BR by 10\%, to explore the
  effects on \ratio.}  The results are summarized in
Table~\ref{tab:model}. {The abundances vary at most by factors of a
  few. Noticeable is the increase in $n(\ce{NH})$  associated with the opening of the NH channel of the DR of \ce{N2H+}. In any
  case, \ce{NH} remains underabundant by almost two orders of
  magnitude. This deficit } is reminiscent of the early results
obtained in the diffuse medium that motivated the {recourse to
  surface reactions. We recall that our models do not include
  freeze-out onto dust grains to enhance the role of gas-phase
  reactions. In the case of NH formation, however, the role of surface
  reactions is poorly constrained.}

\begin{table}
  \begin{center}
    \caption{Predicted steady-state fractional abundances of several
      nitrogen-bearing species in three different models.}
    \scriptsize
    \begin{tabular}{l l l l c}
      \hline
      \hline
      Species   & I & II & III & Observations \\
      \hline
      \ce{N2H+}\tfm{a} & 100:0            & 90:10            & 90:10\\
      \ce{NH4+}\tfm{b} & 85:2:13          & 85:2:13          & 95:2:3\\
      \hline
      \ce{N2}   & $1.0( -5)$ & $1.0( -5)$ & $1.0( -5)$ & $-$ \\
      \ce{N}    & $2.4( -7)$ & $2.5( -7)$ & $2.9( -7)$ & $-$ \\
      \ce{NH}   & $3.5(-11)$ & $9.5(-11)$ & $8.8(-11)$ & $2.0\pm1.0(-9)$ \\
      \ce{NH2}  & $1.9(-10)$ & $1.9(-10)$ & $1.3(-10)$ & $4.0\pm2.0(-10)$ \\
      \ce{NH3}  & $3.7( -8)$ & $3.7( -8)$ & $8.3( -8)$ & $0.3^{+0.1}_{-0.1}-2^{+0.8}_{-0.8}(-8)$ \\
      \hline    
      \ce{NH}:\ce{NH2}:\ce{NH3} & 0.2:1:190 & 0.5:1:190 & 0.7:1:640 & 5:1:300\\  
      \hline
    \end{tabular}
  \tfoot{Numbers in parenthesis are powers of ten. {Branching
      ratios (BR) for the dissociative recombination (DR) of \ce{NH2+}
      are \ce{N + H2} (4\%), \ce{NH + H} (39\%) and \ce{N + 2H} (57\%)
      \citep{thomas2005}.}  \tft{a}{BR for the DR of \ce{N2H+} into
      \ce{N2 + H} and \ce{NH + N}, respectively
      \citep{molek2007,adams2009}.}  \tft{b}{BR for the DR of
      \ce{NH4+} into \ce{NH3 + H}, \ce{NH2 + H2}, and \ce{NH2 + 2H},
      respectively \citep{ojekull2004}.  See additional details in
      Table~\ref{tab:dr}.}}
  \label{tab:model}
  \end{center}
\end{table}

\section{Conclusions}

We have presented {absorption spectra} of the hyperfine structure of
\ce{NH2} and several transitions of \ce{NH3}. These lines emanate from
the low density envelope of the protostar, at densities lower than
typically \dix{6}~\ccc, representative of typical dark cloud
conditions. We have determined the column densities, as well as fractional
abundances. We have found that \ratio$\approx$5:1:300. On the basis of on an updated
chemical network, we have computed the steady-state abundances of amidogen
and ammonia, which agree well with observed values. Imidogen is
undepredicted by more than one order of magnitude, and
$\ce{NH}/\ce{NH2}\le 1$. At this point, more {observations (of
  \eg\ $^{15}$N isotopologues)} and modelling are clearly
needed. Freeze-out and surface reactions, not included in our
calculations, are potentially important. {Before having recourse
  to dust surface processes, however, gas-phase chemistry has to be
  explored in far more details in studying \eg, the influence of the
  gas-phase C/O ratio \citep{hilyblant2010n} or the effects of the
  \ce{H2} o:p ratio}.  The consequences of the uncertainties in the
rates of the dominant chemical paths (dissociative recombinations,
neutral-neutral reactions at low temperature) shall also be explored
in the process.

\begin{acknowledgements}
  We thank the anonymous referee for useful comments. This paper
  benefitted from the CDMS and JPL databases. A.~Faure is warmly
  acknowledged for his careful review of the reaction rates. HIFI has
  been designed and built by a consortium of institutes and university
  departments from across Europe, Canada and the United States under
  the leadership of SRON Netherlands Institute for Space Research,
  with major contributions from Germany, France and the US. Consortium
  members are: Canada: CSA, U.Waterloo; France: CESR, LAB, LERMA,
  IRAM; Germany: KOSMA, MPIfR, MPS; Ireland, NUI Maynooth; Italy: ASI,
  IFSI-INAF, Osservatorio Astrofisico di Arcetri-INAF; Netherlands:
  SRON, TUD; Poland: CAMK, CBK; Spain: Observatorio Astron\'omico
  Nacional (IGN), Centro de Astrobiolog\'{\i}a (CSIC-INTA). Sweden:
  Chalmers University of Technology - MC2, RSS \& GARD; Onsala Space
  Observatory; Swedish National Space Board, Stockholm University -
  Stockholm Observatory; Switzerland: ETH Zurich, FHNW; USA: Caltech,
  JPL, NHSC.
\end{acknowledgements}

\bibliographystyle{aa}
\bibliography{general,cores,phb_refereed,technic,disks}

\Online

\begin{appendix}


\section{Determination of the column density}
\label{sec:cdens}

\begin{figure}
  \begin{center}
    \includegraphics[height=0.8\hsize,angle=-90]{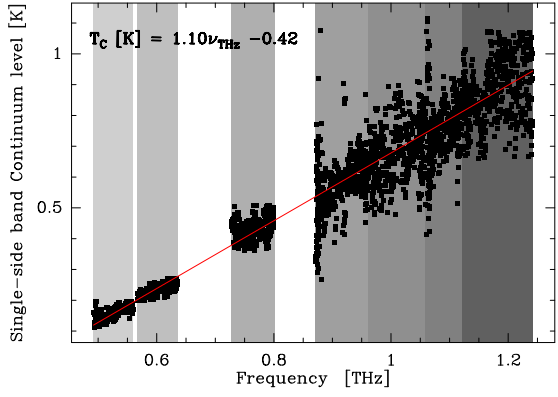}
    \caption{Double sideband continuum (in antenna temperature scale)
      level towards \iras, from 0.5 to 1.2~THz. The straight line is
      the result of a linear fit. The observed HIFI bands are
      highlighted.}
    \label{fig:contlevel}
  \end{center}
\end{figure}

Figure~\ref{fig:contlevel} displays the double-sideband continuum
intensity measured from 0.5 to 1.2~THz with the double-sideband
receivers of the HIFI instrument. The single-sideband continuum is
then estimated assuming equal image and signal gains. The SSB
continuum intensity was estimated following two independent
methods. First, the median of the intensity in each 1~GHz sub-band was
computed. The resulting double-sideband (DSB) continuum level \tc,
sampled every 0.25~GHz from 0.492~THz to 1.242~THz, was found to
increase linearly with the frequency (see
Fig.~\ref{fig:contlevel}). The second method consisted in deconvolving
the despured spectra prior to baseline subtraction, the result of
which is the SSB continuum level. The DSB deconvolution was applied to
the subset of spectra covering the spectral ranges of the lines
considered in this paper. The DSB continuum level is twice the SSB to
better than 10\% in these intervals. From the first method, the
antenna temperature scale SSB continuum level is well fitted by a 1st
order polynomial as $\tc {\rm [K]} = 1.10\,\nu_{\rm THz}-0.42$, for
rest frequencies $\nu_{\rm THz}$ ranging from 0.492 to 1.242~THz (see
Fig.~\ref{fig:contlevel}). The increase in \tc\ with frequency might
be caused by the convolution of the dust temperature profile with the
telescope beam. It may also trace the increase in the dust emissivity
with frequency.  At the higher frequencies of the ammonia lines in
Band~7, the continuum level was estimated using the second
method, in HIPE.

\begin{figure}
  \begin{center}
    \includegraphics[height=0.8\hsize,angle=-90]{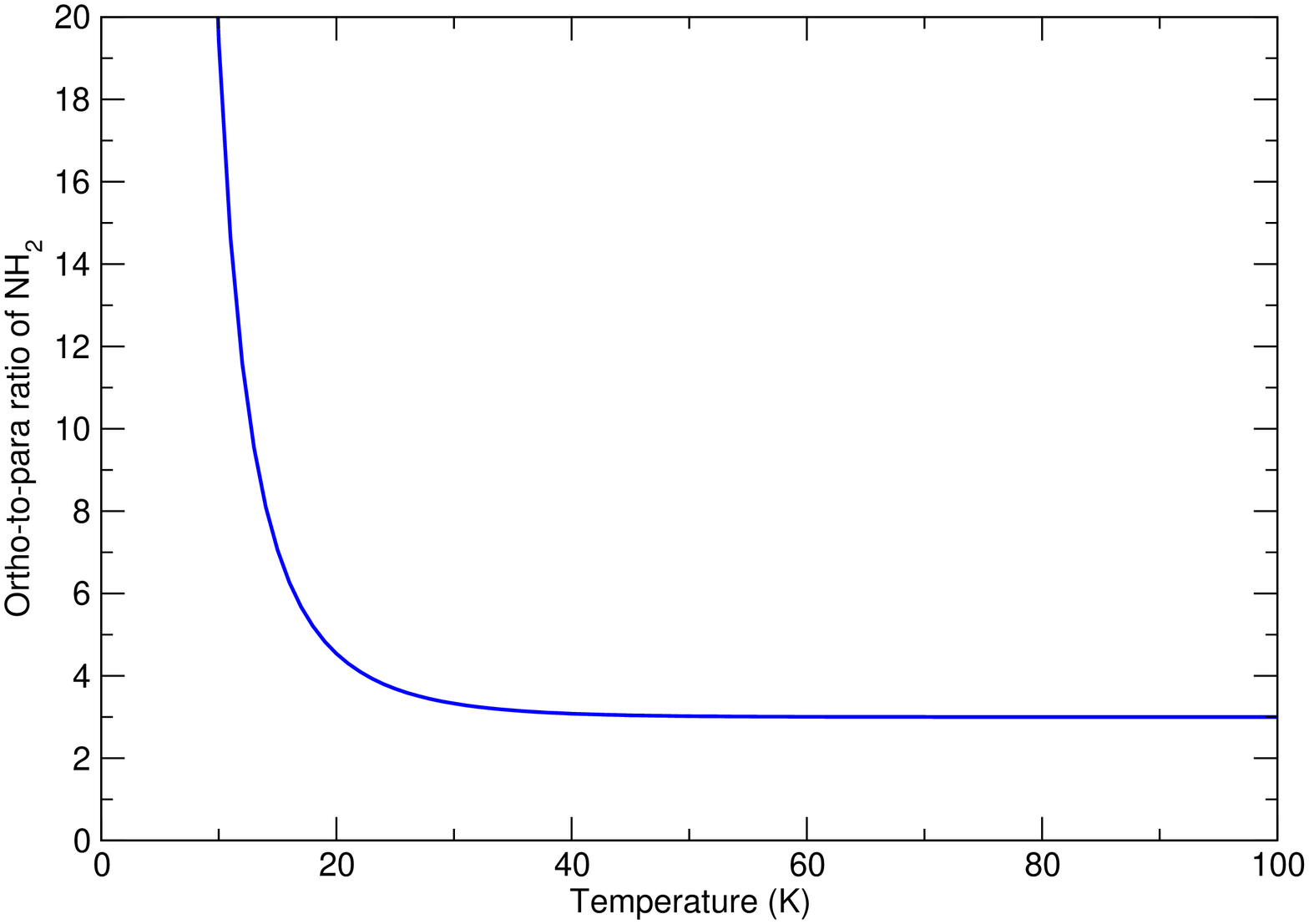}
    \caption{Thermalized ortho:para ratio for \ce{NH2}.}
    \label{fig:opnh2}
  \end{center}
\end{figure}

Figure~\ref{fig:opnh2} shows the o:p ratio for \ce{NH2}
assuming equilibrium at a single temperature for all levels.
\begin{figure}
  \begin{center}
    \includegraphics[height=0.8\hsize,angle=-90]{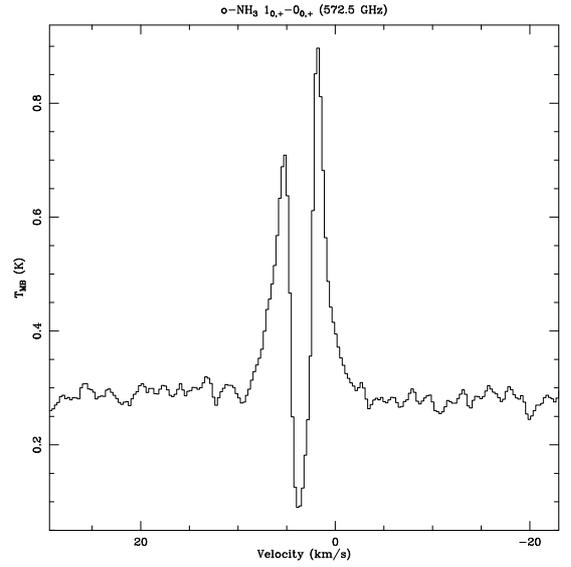}
    \caption{Emission line of the fundamental transition of \ce{NH3}
      at 572.6~GHz detected by HIFI towards \iras.}
    \label{fig:nh3emiss}
  \end{center}
\end{figure}


\section{Chemical modelling}
\label{sec:app-model}

Nitrogen chemistry starts with the formation of {NO and CN by means of
  the reactions $\rm N(OH,NO)H$ and $\rm N(CH,H)CN$, which then lead
  to \ce{N2} \citep{hilyblant2010n}.} Once \nn\ is formed, it reacts
with $\rm He^+$ to form $\rm N^+$, which, by successive {hydrogen
  abstractions}, leads quickly to \ce{NH+}, \ce{NH2+}, \ce{NH3+}, and
\ce{NH4+}. {The dissociative recombination (DR) of \ce{NH4+} is the
  dominant formation route for \ce{NH2} and \ce{NH3}. The formation of
  NH is dominated by \ce{NH2}(O,H)NH, unless the NH channel of the DR
  of \ce{N2H+} has a non-zero branching ratio (BR). Although DR
  branching ratios have been (re)measured recently for \ce{N2H+},
  \ce{NH2+}, and \ce{NH4+}, there are still disparate results and
  significant uncertainties. According to the most recent litterature
  \citep[see Table~\ref{tab:model}\,and][]{florescu2006}, the BR of DR
  reactions are uncertain by at least 10\%.  We conducted three model
  calculations, where we varied the BR by 10\%, to explore the effects
  on \ratio.}  The results are summarized in
Table~\ref{tab:model}. {The abundances vary at most by factors of a
  few. We note the increase in $n(\ce{NH})$, which is associated with
  the opening of the NH channel of the DR of \ce{N2H+}. In any case,
  \ce{NH} remains underabundant by almost two orders of
  magnitude. This deficit } is reminiscent of the early results
obtained in the diffuse medium that motivated the {recourse to surface
  reactions. We recall that our models do not include freeze-out onto
  dust grains to ensure that the role of gas-phase reactions is
  enhanced. In the case of NH formation, however, the role of surface
  reactions is largely ill-constrained.}

To model the steady-state abundances of the nitrogen hydrides, the
\osu\ network and rates were used.  The \osu\ network contains 13
elements, 449 species, and 4457 gas-phase reactions. We note that this
version of the OSU database does not contain molecular anions or any
depletion of gas-phase species. The network has been updated from a
revision of the branching ratios and rate coefficients for the
dissociative recombination (DR) of the nitrogen bearing cations
\ce{N2H+}, \ce{NH2+}, and \ce{NH4+}:
\begin{itemize}
\item The DR of \ce{N2H+} has been determined experimentally using
  both flowing afterglow \citep[FA, see ][]{adams1991} and storage
  ring (SR) techniques \citep{geppert2004}, leading to controversial
  results concerning the branching ratios (BR) of the two channels
  \ce{N2 + H} and \ce{NH + N}. Thus, in contrast to the FA results that
  established the major product as \ce{N2 + H} with a BR $\approx
  100$\%, \citet{geppert2004} found this channel to account for only
  36\% of the total reaction. The most recent FA and SR measurements
  \citep{molek2007, adams2009}, however, have confirmed the earlier FA
  results that the DR of \ce{N2H+} should lead predominantly to \ce{N2
    + H} with a BR $\approx 90-100$\%. This result is also supported
  by the {\it ab initio} calculations of \citet{talbi2009}. For the
  total rate coefficient, we adopted the (temperature dependent)
  expression of \citet{geppert2004}, as in the \osu\ network.
\item For the DR of \ce{NH2+}, the latest SR measurements are those
  of \citet{thomas2005}, who obtained the BR for \ce{N + H2}
  (4\%), \ce{NH + H} (39\%), and \ce{N + 2H} (57\%). For the rate
  coefficient, we adopted the expression recommended by
  \citet{mitchell1990}, as in the \osu\ network.
\item Finally, the SR measurements of \citet{ojekull2004} demonstrated
  that the DR of \ce{NH4+} is dominated by the product channels
  \ce{NH3 + H} (85\%), \ce{NH2 + 2H} (13\%), and \ce{NH2 + H2}
  (2\%). For the total rate, we adopted the expression of
  \citet{ojekull2004}, which differs slightlty from the one
  recommended in the \osu\ network.
\end{itemize}
We note that for \ce{NH3+} there is to our knowledge neither
measurements nor calculations available. We therefore adopted the rate
and branching ratios recommended in the \osu\ network, corresponding
to the two channels \ce{NH + 2H} (50\%) and \ce{NH2 + H} (50\%).

\begin{figure}
  \centering
  \includegraphics[width=0.8\hsize,angle=0]{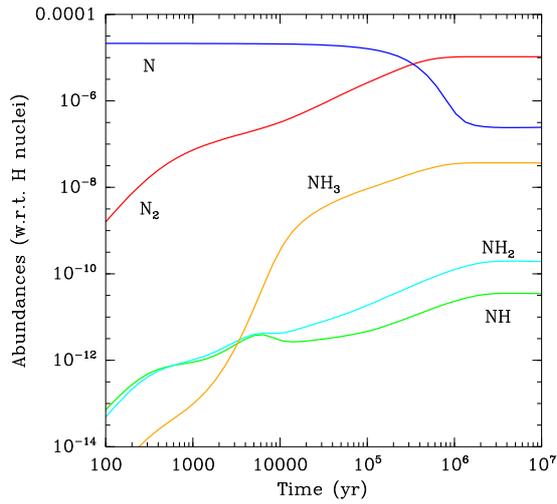}
  \caption{Predicted abundances for several N-bearing molecules as a
    function of time for model 1 ("standard", see text), for a density
    $\nh=2\tdix{4}$~\ccc, a gas temperature \tkin=10~K, and a total
    extinction \av=10~mag. Abundances are given with respect to H
    nuclei.}
  \label{fig:model}
\end{figure}

Following the above update of DR rates and branching ratios, we have
adopted three different chemical models by varying the DR branching
ratios of \ce{N2H+} and \ce{NH4+} within 10\%, the typical
experimental uncertainty. The DR of these two ions was indeed found
to be the dominant formation routes of NH and \ce{NH2},
respectively. The employed branching ratios are listed in
Table~\ref{tab:dr}.

\begin{table}
  \caption{Dissociative recombination branching ratios employed in the
    present work.}  \centering
  \centering
  \begin{tabular}{l c c l r r r c}
    \hline
    \hline
    Ion & \multicolumn{2}{c}{Total rate\tfm{a}} & Products & \multicolumn{3}{c}{Model} & \\ 
        & $\gamma$ & $\beta$ & & 1 & 2 & 3 \\\hline
    \ce{N2H+}\tfm{b} & 1.00(-7) & -0.5   & \ce{N2 + H}      & 100\% &  90\% &  90\%\\
              & & & \ce{NH + H}  & 0\%   &  10\% &  10\% \\
    \ce{NH2+}\tfm{c} & 3.00(-7) & -0.5   & \ce{N + H + H}& 57\% &  &   \\
              &                & &\ce{N + H2}      &  4\% &  &     \\
              &                & &\ce{NH + H}      & 39\% &  &     \\
    \ce{NH3+} & 3.10(-7) & -0.5 & \ce{NH  + H + H} & 50\% \\
              & &            & \ce{NH2 + H}     & 50\% \\
    \ce{NH4+}\tfm{d} & 9.40(-7) & -0.6   & \ce{NH3 + H}     & 85\%  &  85\% &  95\%  \\
              & & & \ce{NH2 + H2}    &  2\%  &   2\% &   2\% \\
              & & & \ce{NH2 + H + H} & 13\%  &  13\% &   3\% \\
    \hline
  \end{tabular}
  \label{tab:dr}
  \tfoot{\tft{a}{Reaction rates are written as $k=\gamma(T/300)^\beta$
      in cm$^3$\s.}
    \tft{b}{\cite{molek2007}}\tft{c}{\cite{thomas2005}}
    \tft{d}{\cite{ojekull2004}}.}
\end{table}

\end{appendix}
\end{document}